\documentclass[twocolumn,pre,superscriptaddress,aps,footinbib]{revtex4}
\usepackage{url,psfrag,graphicx}
\usepackage{dcolumn}
\usepackage{amsmath,amssymb}
\usepackage{bm}
\usepackage{hyperref}
\usepackage{float,epsfig,color}
\usepackage{times}
\usepackage[latin1]{inputenc}

\usepackage{graphicx}
\usepackage{amsmath}
\usepackage{amssymb}
\usepackage{dcolumn}

\begin{document}

\title{Transition between chaotic and stochastic universality classes of kinetic roughening}

\author{Enrique Rodr\'{\i}guez-Fern\'andez}
\email{enrodrig@math.uc3m.es}
\affiliation{Departamento de Matem\'aticas and Grupo Interdisciplinar de Sistemas Complejos (GISC)\\ Universidad Carlos III de Madrid, Avenida de la Universidad 30, 28911 Legan\'es, Spain}
\author{Rodolfo Cuerno}
\email{cuerno@math.uc3m.es}
\affiliation{Departamento de Matem\'aticas and Grupo Interdisciplinar de Sistemas Complejos (GISC)\\ Universidad Carlos III de Madrid, Avenida de la Universidad 30, 28911 Legan\'es, Spain}

\begin{abstract}
The dynamics of non-equilibrium spatially extended systems are often dominated by fluctuations, due to e.g.\ deterministic chaos or to intrinsic stochasticity. This reflects into generic scale invariant or kinetic roughening behavior that can be classified into universality classes defined by critical exponent values and by the probability distribution function (PDF) of field fluctuations. Suitable geometrical constraints are known to change secondary features of the PDF while keeping the values of the exponents unchanged, inducing universality subclasses. Working on the Kuramoto-Sivashinsky equation as a paradigm of spatiotemporal chaos, we show that the physical nature of the prevailing fluctuations (chaotic or stochastic) can also change the universality class while respecting the exponent values, as the PDF is substantially altered. This transition takes place at a non-zero value of the stochastic noise amplitude and may be suitable for experimental verification.
\end{abstract}

\maketitle

Generic scale invariance (GSI) describes the behavior of many spatially extended non-equilibrium systems in which driving and dissipation act at comparable rates, such that strong correlations build up whose space-time behavior lacks characteristic scales \cite{Grinstein95}. Hence, they are analogous to equilibrium critical systems, a crucial difference being that parameter tuning is not required for criticality in GSI \cite{Sethna06}. Thus, GSI is one of the forms in which critical dynamics \cite{Taeuber14} is being recently generalized to novel non-equilibrium contexts as assessed in physical and non-physical systems, from quantum matter  \cite{Sieberer13,Sieberer18}, to living \cite{Munoz18}, or social \cite{Perc17} systems.

A key player in recent advances on the understanding of driven systems displaying GSI
\cite{Barabasi95,Krug97,Kriechebauer10,Halpin-Healy15,Takeuchi18} is the celebrated Kardar-Parisi-Zhang (KPZ) equation for the height $h(x,t)$ of an interface at a substrate position $x\in\mathbb{R}^d$ and time $t$ \cite{Kardar86}, subject to fluctuations, namely
(we henceforth set $d=1$),
\begin{eqnarray}
 & \partial_t h = \nu \partial_x^2 h + \dfrac{\lambda}{2} (\partial_x h)^2 + D \, \eta , & \label{eq:kpz} \\
 & \langle \eta(x,t)\eta(x',t') \rangle = \delta (x-x') \delta (t-t') , &
\label{eq:noise_kpz}
\end{eqnarray}
where $\nu, D >0$, and $\lambda$ are parameters, and $\eta(x,t)$ is zero-average, Gaussian noise.
Within the physical image of evolving interfaces, the GSI behavior displayed by the KPZ equation and related stochastic systems is termed kinetic roughening \cite{Barabasi95,Krug97}. In analogy with equilibrium critical dynamics, it can be classified into universality classes, which are determined by the values of critical exponents and by the probability distribution function (PDF) of, say, height fluctuations \cite{Kriechebauer10,Halpin-Healy15,Takeuchi18}. Indeed, the 1D KPZ universality class is recently being identified in the fluctuation dynamics of a wide range of low-dimensional, strongly-correlated systems, from thin-film growth, random polymers, or randomly-stirred fluids \cite{Barabasi95,Krug97}, to active matter \cite{Chen16}, quantum entanglement \cite{Nahum17}, or spatiotemporal chaos \cite{Roy20}, to cite a few.

Based on exact solutions of the one-dimensional (1D) KPZ equation, Eq.\ \eqref{eq:kpz}, and related discrete models (see \cite{Kriechebauer10,Halpin-Healy15,Takeuchi18} for reviews), a particularly rich structure is being elucidated for GSI universality classes. To begin with, critical exponent values are known {\em not} to unambiguously identify a given class. Indeed, explicit examples have been reported of {\em linear} height equations with time- \cite{Saito12} or space- \cite{Rodriguez-Fernandez20} correlated noise, which feature the 1D KPZ scaling exponents, but which cannot be in this universality class, their field PDF being Gaussian, while for the {\em nonlinear} 1D KPZ equation it is a member of the Tracy-Widom (TW) PDF family
\cite{Kriechebauer10,Halpin-Healy15,Takeuchi18}.

Moreover, while keeping the same values of the scaling exponents, universality {\em subclasses} exist of the 1D KPZ class, which differ by the flavor of the precise TW PDF which occurs: e.g.\ for globally flat (curved) interfaces growing from a straight line (point), it corresponds to the largest-eigenvalue TW distribution of random matrices in the Gaussian orthogonal (unitary) ensemble [GOE (GUE)] \cite{Kriechebauer10,Halpin-Healy15,Takeuchi18}. Equivalently, finite systems whose size decreases (increases) linearly with time display TW-GOE (GUE) statistics \cite{Carrasco14,Fukai17,Carrasco18}, while analogous transitions have been assessed for changes in the background topology \cite{Santalla17} or in the rate of system-size change \cite{Carrasco19}. Furthermore, the existence of universality subclasses induced by similar changes in geometrical constraints carries over to the main linear \cite{Carrasco19b} and nonlinear \cite{Carrasco16} universality classes of kinetic roughening other than KPZ, making this a robust trait of (this type of) criticality far from equilibrium.

However, in all these cases the universality subclasses sharing scaling exponent values differ by some global geometrical or topological condition on the system size or background metric, the existence of alternative mechanisms which likewise control the field PDF remaining uncertain. In this Letter, we demonstrate the physical nature of the prevailing system fluctuations as one such mechanism. Specifically, by considering the 1D Kuramoto-Sivashinsky (KS) equation \cite{Kuramoto76,Sivashinsky77} for a scalar field $u(x,t)$, which reads
\begin{equation}
\partial_t u = -\nu_0 \partial_x^2 u -\kappa_0 \partial_x^4 u + \lambda_0 u \partial_x u + (D_0 + \tilde{D}_0 \partial_x) \eta , 
\label{eq:kuramoto}
\end{equation}
where $\nu_0, \kappa_0, D_0, \tilde{D}_0>0$, and $\lambda_0$ are parameters, and $\eta$ is as in Eq.\ \eqref{eq:noise_kpz}, we show (for $D_0=0$, $\tilde{D}_0\neq 0$) that the kinetic roughening behavior which it displays is characterized by critical exponent values which are noise-independent, while the field PDF is non-Gaussian (Gaussian) for low (large) noise values, corresponding to dynamics dominated by chaotic (stochastic) fluctuations. This
transition in the universality class occurs at a non-zero noise amplitude $\tilde{D}_0$ and might be observable in suitable experimental contexts.

The {\em deterministic} ($D_0=\tilde{D}_0=0$) KS equation is a paradigm of spatiotemporal chaos (SC) \cite{Cross93,Cross94}, where it
has become a benchmark to assess novel concepts and tools, like e.g.\ reservoir computing \cite{Pathak17,Pathak18}. Either Eq.\ \eqref{eq:kuramoto} proper or the equally ubiquitous version of the deterministic KS equation \cite{Kuramoto76,Sivashinsky77}, satisfied by $h(x,t)=\int_0^x u(y,t) {\rm d}y$ [see Eq.\ \eqref{eq:kuramotoKPZ} below], both provide physical models in many different contexts, from liquid flow down inclines \cite{Nepomnyashchii74,Pradas12} to solidification \cite{Misbah91}. The {\em stochastic} equation for $h$ has been derived in e.g.\ epitaxial growth \cite{Karma93,Misbah10}, ion-beam sputtering \cite{Cuerno95,Lauritsen96}, or diffusion-limited growth \cite{Cuerno01}; in Appendix A we derive Eq.\ \eqref{eq:kuramoto} for a falling liquid film under thermal fluctuations \cite{Seeman01,Mecke05}.

The large-scale behavior of the deterministic KS equation, Eq.\ \eqref{eq:kuramoto}, is known to remarkably coincide \cite{Yakhot81,Hayot93} with that of the {\em stochastic} Burgers equation, whose GSI exponents \cite{Forster77} and field PDF \cite{Rodriguez-Fernandez19,Rodriguez-Fernandez20} are known. Likewise, the deterministic \cite{Sneppen92,Roy20}, as well as the
stochastic KS equations for $h$ \cite{Cuerno95b,Ueno05} are both in the KPZ universality class. However, how and if the nature of the fluctuations, whether deterministic chaos or stochastic noise, reflects into the GSI behavior, has remained overlooked thus far.

We begin by investigating in full detail the universality class of Eq.\ \eqref{eq:kuramoto} for the deterministic case, as well as for the stochastic cases with conserved ($D_0=0$, $\tilde{D}_0\neq0$) and non-conserved ($D_0\neq0$, $\tilde{D}_0=0$) noise. Note that, as seminally argued for by Yakhot \cite{Yakhot81}, Eq.\ \eqref{eq:kuramoto} is expected to renormalize at large scales into an effective stochastic Burgers equation,
\begin{equation}
\partial_t u = \nu \partial_x^2 u + \lambda u \partial_x u + (D + \tilde{D} \partial_x) \eta,  \label{eq:burgers}
\end{equation}
where notably $\nu>0,$ rendering asymptotically irrelevant the biharmonic term in Eq.\ \eqref{eq:kuramoto}. Moreover, the noise in the effective equation, Eq.\ \eqref{eq:burgers}, respects the conservation law expressed by Eq.\ \eqref{eq:kuramoto}. I.e, if the bare equation is deterministic or has conserved noise, then $D=0$, $\tilde{D}\neq0$, while if the bare noise is non-conserved, so is the effective noise, thus $D\neq0$, $\tilde{D}=0$.

We have performed numerical simulations of Eq.\ \eqref{eq:kuramoto} using the pseudospectral method in \cite{Gallego11} and periodic boundary conditions for system size $L=2048$. Initial conditions are random (with $10^{-5}$ amplitude) for the deterministic case and zero otherwise. Parameters are fixed to $\nu_0=\kappa_0=1$, $\lambda_0=10$, and space-time discretization steps $\delta x=1$, and $\delta t \in [0.01,0.05]$.
\begin{figure}[!t]
\begin{center}
\includegraphics[width=1.0\columnwidth]{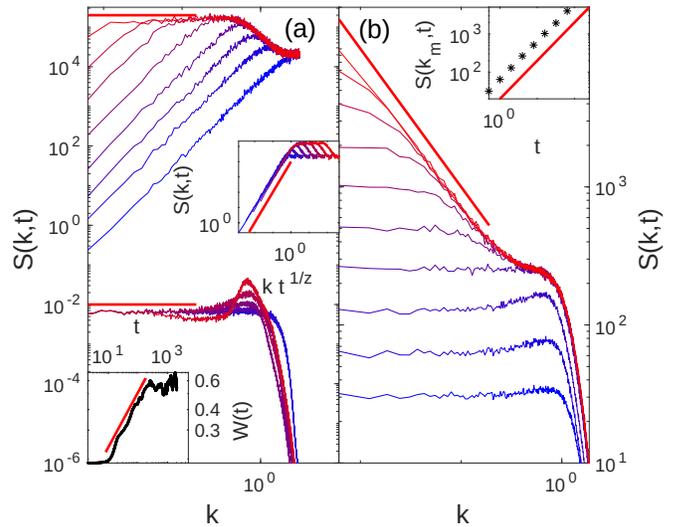}
\caption{Time evolution of $S(k,t)$ from numerical simulations of Eq.\ \eqref{eq:kuramoto} for different noise conditions: (a) deterministic case ($D_0=\tilde{D}_0=0$; bottom) and conserved-noise ($D_0=0,\tilde{D}_0=1$; top). Top inset: data collapse for $\alpha_{\rm cn}=-1/2,z_{\rm cn}=3/2$. Bottom inset: time evolution of the roughness of $h(x,t)=\int_0^x u(y,t) {\rm d}y$; the straight line has slope $\beta_{\rm nc}=(\alpha_{\rm cn}+1)/z_{\rm cn}=1/3$; (b) non-conserved noise ($D_0=1,\tilde{D}_0=0$). Inset: time evolution of $S(k_m,t)$ for the smallest $k$-value, $k_m$; the straight line has slope $2\alpha_{\rm nc}+1=z_{\rm nc}=1$. For all panels, averages are over $100$ realizations, time increases from blue to red, and the slope of the red solid line for small $k$ is $-(2\alpha+1)$ for $\alpha=\alpha_{\rm cn}=-1/2$ [$\alpha_{\rm nc}=0$] on (a) [(b)].}
\label{fig:Sk}
\end{center}
\end{figure}
Under GSI conditions \cite{Barabasi95,Krug97}, the scaling exponents characterizing the universality class can be readily identified in the evolution of the field roughness $W$ (rms deviation of the field fluctuations), which increases with time as $W \sim t^{\beta}$, reaching a saturation value $W_{sat} \sim L^{\alpha}$ at time $t_{sat} \sim L^z$, where $z=\alpha\beta$ is the dynamic exponent \cite{Taeuber14} and $\alpha$ is the roughness exponent, related with the fractal dimension of the $u(x)$ profile \cite{Barabasi95}. The same exponents occur in the two-point statistics, e.g.\ in the structure factor $S(k,t)=\langle |\hat{u}(k,t)|^2 \rangle$, where hat denotes space Fourier transform and $k$ is wavenumber (two-point correlations in real space are similarly assessed in Appendix A). Indeed \cite{Barabasi95,Krug97}, $S(k,t)\sim 1/k^{2\alpha+d}$ for $t\gg L^z$, while $S(k,t) \sim t^{(2\alpha+d)/z}$ for $k \ll 1$ (with $d=1$ here). The numerical time evolution of $S(k,t)$ for the various noise conditions is shown in Fig.\ \ref{fig:Sk}, where the scaling exponents predicted by Yakhot's argument \cite{Yakhot81} are indeed obtained: both in the deterministic and in the conserved-noise cases, Eq.\ \eqref{eq:kuramoto} renormalizes into Burgers equation, Eq.\ \eqref{eq:burgers}, with conserved noise $D=0, \tilde{D}\neq0$, for which $\alpha_{\rm cn}=-1/2$ and $z_{\rm cn}=3/2$ \cite{Rodriguez-Fernandez20}. While $S(k,t)$ approaches the ($k$-independent) white-noise behavior for increasing $t$ in these two cases, the detailed form of the structure factor curves differs noticeably. In the deterministic case, the scaling exponents are obtained via the integrated $h$ field, trivially expected to scale as $h \sim x^{\alpha+1}$ \cite{Kardar86,Rodriguez-Fernandez20}. In contrast, for non-conserved noise, Eq.\ \eqref{eq:kuramoto} now renormalizes into a noisy Burgers equation, Eq.\ \eqref{eq:burgers}, with non-conserved noise $D\neq 0, \tilde{D}=0$, for which $\alpha_{\rm nc}=0$ and $z_{\rm nc}=1$ \cite{Rodriguez-Fernandez19}.

\begin{figure}[t!]
\begin{center}
\includegraphics[width=1.0\columnwidth]{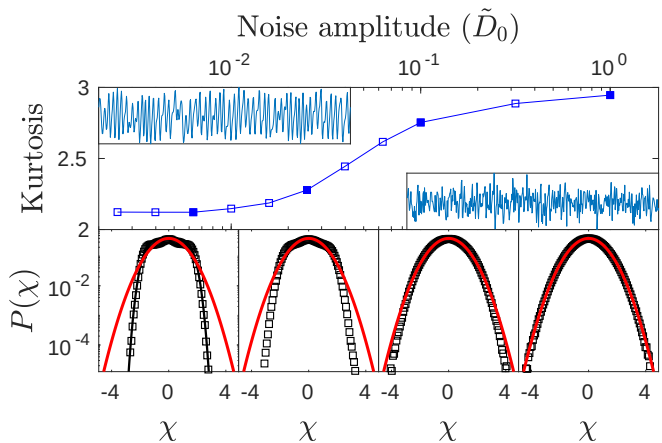}
\caption{Top panel: Kurtosis of $u$-fluctuations within the nonlinear regime prior to saturation for Eq.\ \eqref{eq:kuramoto} with $D_0=0$, and different values of $\tilde{D}_0$. The line is a guide to the eye. The bottom panels show the PDF of standardised $u$-fluctuations ($\chi$) for increasing values of $\tilde{D}_0$ left to right, which correspond to the filled squares in the top panel. Red solid lines show an exact Gaussian PDF; the black solid line is for $P[\chi] \sim \exp (-\chi^{4.5})$. Insets in top panel show representative $u(x)$ profiles for $\tilde{D}_0=0$ (deterministic case; left) and $\tilde{D}_0=1$ (conserved-noise case; right). Averages are over $10$ realizations.}
\label{fig:PDFs}
\end{center}
\end{figure}

The scaling exponent values already imply that the KS equation with non-conserved noise belongs to a different universality class than the deterministic and conserved-noise equations, which share the same values of $\alpha$ and $z$. Hence, for the remainder of this work we set $D_0=0$ and focus on the latter two cases. Still, as noted above, the values of the two independent scaling exponent are currently known \cite{Saito12,Rodriguez-Fernandez20} {\em not} to necessarily fix the GSI universality class unambiguously.
Moreover, the 1D KPZ universality class also illustrates the fact \cite{Kriechebauer10,Halpin-Healy15,Takeuchi18} that the field PDF can differ in the nonlinear regime prior to saturation to a steady state, as compared with the PDF of fluctuations around such a steady state after it has been reached. Hence, we next investigate the fluctuation statistics for the $u$ field in Eq.\ \eqref{eq:kuramoto} within the nonlinear regime prior to saturation, setting $D_0=0$ and considering different values of the conserved-noise amplitude $\tilde{D}_0$; results are shown in Fig.\ \ref{fig:PDFs}.
In the deterministic $\tilde{D}_0=0$ case, the rescaled fluctuations of $u$ around its space average $\Bar{u}$, defined as $\chi=(u-\Bar{u})/\text{std}(u)$, exhibit a symmetric probability density function (PDF) whose tails decay much faster than those of a Gaussian distribution. Hence, the kurtosis is much smaller than 3, see Fig.\ \ref{fig:PDFs}. This PDF features two symmetric shoulders implying a relatively high frequency for two characteristic fluctuations in $u$-values, which can be approximately identified by inspection of the $u(x)$ profile shown in the figure. Similar distributions had been earlier reported at steady state \cite{Hayot93}. We assess the full time evolution of the skewness and the excess kurtosis of the $u$ fluctuations in Appendix A, finding them to remain virtually unchanged along the nonlinear time regime. The main qualitative features of the PDF are preserved for increasing values of the conserved-noise amplitude $\tilde{D}_0$, up to a certain value. For larger values of $\tilde{D}_0$, the fluctuation PDF starts to approach the Gaussian form, with a kurtosis which approaches the exact Gaussian value, see Fig.\ \ref{fig:PDFs}. Inspection of the representative $u(x)$ profile shown for $\tilde{D}_0=1$ indeed suggests the smaller predominance of characteristic fluctuations around the mean than in the deterministic case. Further details on the transition of the PDF with $\tilde{D}_0$ are provided in Appendix A.

While the PDFs of the deterministic and the (large) conserved-noise cases of Eq.\ \eqref{eq:kuramoto} are both even in $\chi$, they are obviously different, specially with respect to the occurrence of ``typical'' fluctuation values. One could speak of two different subclasses of a single universality class which additionally features $\alpha_{\rm cn}=-1/2$ (white noise) and $z_{\rm cn}=3/2$ (superdiffusive spread of correlations, as in the 1D KPZ equation). However, further dynamical properties suggest that we rather speak of a change in the universality class, with a transition at a well-defined value of $\tilde{D}_0$ separating the predominance of chaotic or of stochastic fluctuations. This interpretation is supported by a study of the behavior of the finite-size Lyapunov exponents of the system \cite{Pikovsky16,Cencini13} as a function of the conserved-noise amplitude. Specifically, we measure the so-called scale-dependent Lyapunov exponent $\Lambda(\epsilon)$ \cite{Gao06,Cencini13}, where $\epsilon$ is a distance between trajectories $\{T_i(x_0)=[u(x_0,(i+1)\Delta t),u(x_0,(i+2)\Delta t),...,u(x_0,(i+m)\Delta t)]\}_{i=1}^I$ extracted from the time series $u(x_0,t)$ where $x_0$ is any fixed value of $x$ and $\Delta t$ is a sampling time. These trajectories are efficient tools to reconstruct the attractor of the dynamics \cite{Pikovsky16}. Once all $(i,j)$ pairs of close-by trajectories are selected for a certain $\epsilon_0 \ll 1$ such that
$ 0 < || T_i-T_j ||/\max_{k,l} || T_k-T_l || < \epsilon_0 \ll 1, $
we compute
$ \Lambda(\epsilon_t) = \Delta t^{-1} \log{ \epsilon_{t+\Delta t}/\epsilon_t } $
for each selected $(i,j)$ and for several times $t=N\Delta t$, where $ \epsilon_t = || T_{i+N} - T_{j+N} ||$ and $ \epsilon_{t+\Delta t} = || T_{i+N+1} - T_{j+N+1} || $. Results are averaged for different values of $x_0$ and, after binning, an $\epsilon$-dependent Lyapunov exponent $\Lambda(\epsilon)$ is obtained. For purely stochastic systems, $\Lambda(\epsilon)$ has been shown to display monotonic power-law decay with $\epsilon$  \cite{Gao06}; in contrast, chaotic systems are characterized by a well-defined plateau where $\Lambda(\epsilon)$ shows $\epsilon$-independent behavior, for not too large $\epsilon$ values \cite{Gao06}.

\begin{figure}[t!]
\begin{center}
\includegraphics[width=1\columnwidth]{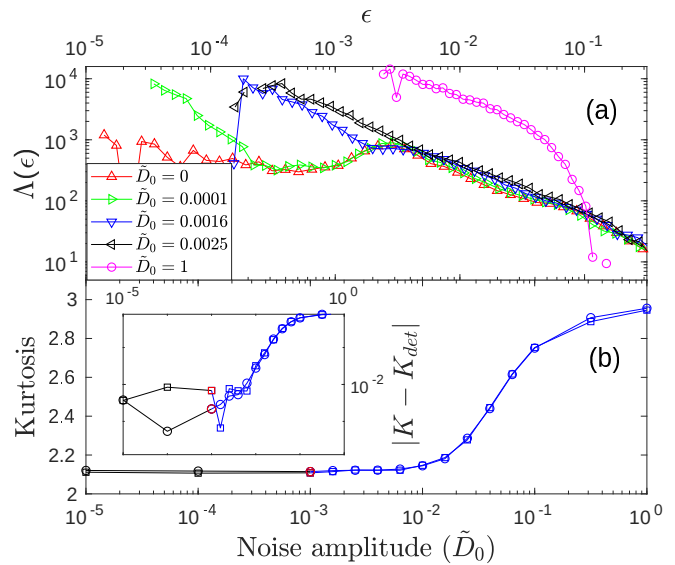}
\caption{(a) Scale-dependent Lyapunov exponent vs $\epsilon$ from numerical solutions of Eq.\ \eqref{eq:kuramoto} within the nonlinear regime prior to saturation, for $D_0=0$ and conserved-noise amplitudes $\tilde{D}_0$ as in the legend. (b) Kurtosis of $u$ fluctuations vs $\tilde{D}_0$ for $L=1024$ (squares) and $L=4096$ (circles). Inset: Distance bewteen $\mathcal{K}(\tilde{D}_0)$ and the kurtosis of the deterministic system, $\mathcal{K}_{det}$, vs $\tilde{D}_0$. For each $L$, the difference grows (blue data) to the right of the corresponding red point. All lines are guides to the eye. Averages are over $10$ realizations.}
\label{fig:LambdaU}
\end{center}
\end{figure}

We have studied numerically the behavior of $\Lambda(\epsilon)$ for Eq.\ \eqref{eq:kuramoto} different values of $\tilde{D}_0$, setting $D_0=0$, $m=4$, $I=4997$, and $\Delta t=\delta t$; results are shown in Fig.\ \ref{fig:LambdaU}.
Indeed, in the $(\tilde{D}_0=0)$ deterministic case $\Lambda(\epsilon)$ displays a well-defined plateau for small $\epsilon$, and decays monotonously for $\epsilon\gtrsim 3 \cdot 10^{-3}$. The plateau width decreases for increasing $\tilde{D}_0$. In contrast, for ``large'' $\tilde{D}_0=1$, $\Lambda(\epsilon)$ decays monotonically with $\epsilon$. We consider that a transition takes place when the plateau first vanishes, for $\tilde{D}_0\simeq \tilde{D}_{0,c}=1.6 \cdot 10^{-3}$. Actually, starting at this value the kurtosis ${\cal K}(\tilde{D}_0)$ departs from its deterministic value, approaching Gaussian behavior. Moreover, the $L$-dependence of the threshold value $\tilde{D}_{0,c}$ seems weak, see Fig.\ \ref{fig:LambdaU}.

\begin{figure}[t!]
\begin{center}
\includegraphics[width=1.0\columnwidth]{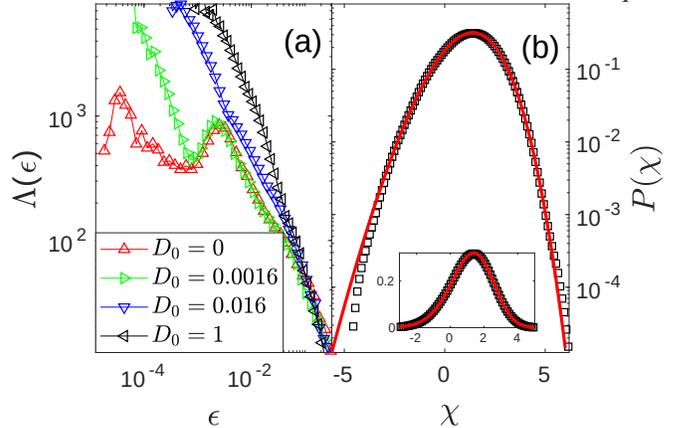}
\caption{(a) Scale-dependent Lyapunov exponent for solutions of the KS Eq.\ \eqref{eq:kuramotoKPZ} within nonlinear regime prior to saturation, for noise amplitudes $D_0$ as in the legend. Remaining parameters as in previous figures. (b) PDF of $h$-fluctuations within nonlinear regime for $D_0=1$. Red solid line shows the exact GOE-TW PDF \cite{Kriechebauer10,Halpin-Healy15,Takeuchi18}. Inset: same data in a linear plot. Averages are over $10$ realizations.}
\label{fig:LambdaH}
\end{center}
\end{figure}

We have additionally assessed this transition between chaotic and stochastic GSI behavior in the KS equation satisfied by the space integral of $u$ defined above, namely,
\begin{equation}
\partial_t h = -\nu_0 \partial_x^2 h - \kappa_0 \partial_x^4 h + \frac{\lambda_0}{2} (\partial_x h)^2 + D_0 \eta ,
\label{eq:kuramotoKPZ}
\end{equation}
where we only consider the non-conserved noise case, relevant within Yakhot's argument \cite{Yakhot81,Sneppen92,Roy20}. In Fig.\ \ref{fig:LambdaH} we confirm that, for an increasing noise amplitude $D_0$, $\Lambda(\epsilon)$ behaves similarly to Eq.\ \eqref{eq:kuramoto}: the ($D_0=0$) deterministic system shows a plateau of $\epsilon$-independent behavior which disappears for $D_0\gtrsim 0.01$, beyond which stochastic behavior ensues. However, now the transition does not reflect into a change between two different fluctuation PDFs. Indeed, in the chaotic case ($D_0=0$) fluctuations of Eq.\ \eqref{eq:kuramotoKPZ} are known to be Tracy-Widom distributed in the nonlinear regime prior to saturation \cite{Roy20}, while Fig.\ \ref{fig:LambdaH} indicates that so do fluctuations for a ``large'' value of stochastic noise. Scaling exponent values are known to be 1D KPZ, independently of the value of $D_0$ \cite{Cuerno95,Roy20}.

In summary, we have seen that Yakhot's classic argument on the asymptotic equivalence between SC and GSI holds only partly for Eqs.\ \eqref{eq:kuramoto} and \eqref{eq:kuramotoKPZ}; specifically, for the conserved KS equation, Eq.\ \eqref{eq:kuramoto}, different universality classes occur with different field PDF, albeit with the same scaling exponent values. This fact is not captured by Yakhot's argument which, while incorporating the basic system symmetries (conserved vs non-conserved dynamics, etc.), misses key differences between PDFs which are otherwise consistent with the former. The dynamical role of symmetries can be subtle indeed in the present class of nonequilibrium critical systems \cite{Sieberer13,Mathey17,Rodriguez-Fernandez19}.

Crucially, each one of the GSI universality classes occurring in Eq.\ \eqref{eq:kuramoto} correlates with the nature (chaotic or stochastic) of the mechanism controlling fluctuations in the system, the transition between them nontrivially occurring at a nonzero stochastic noise amplitude. This transition could be experimentally verified, for instance in epitaxial growth of vicinal surfaces \cite{Misbah10}, where the KS equation describes the dynamics of atomic steps separating terraces under non-negligible adatom desorption \cite{Karma93}. In such a case the equation for the step slope is Eq.\ \eqref{eq:kuramoto}, where $\tilde{D}_0$ scales as an inverse power of the characteristic desorption time.

For the non-conserved KS equation, Eq.\ \eqref{eq:kuramotoKPZ}, although an analogous transition takes place in the dominance of chaotic or stochastic fluctuations, on both sides of the transition the field PDF and the scaling exponent are those of the 1D KPZ universality class. Indeed, the TW distribution is not only relevant to the {\em stochastic} 1D KPZ class, but also describes the fluctuations of {\em deterministic} chaotic systems \cite{Spohn16}. This coincidence might well be accidental and limited to 1D systems. Its exploration in 2D might provide some clue on the relation between the deterministic KS and the stochastic KPZ equations in higher dimensions \cite{Manneville96}, an open challenge in the fundamental understanding of spatiotemporal chaos \cite{Boghosian99}.

With respect to the specifics of the present transition, it would be interesting to obtain analytical estimates on the threshold noise amplitude and to assess nontrivial consequences on physical quantities beyond the field PDF. The behavior discussed above for $S(k,t)$ already indicates differences in the equal-time two-point statistics, but two-time statistics may introduce additional novelties. In this process, it would be interesting to find analogous transitions, but in which the chaotic and stochastic ``phases'' might also differ by the values of the scaling exponents.

Finally, the correlation between the field PDF and the nature of the fluctuations underscores the importance of assessing the PDF explicitly, to correctly identify the GSI universality class. This is particularly critical in view of the plethora of experimental complex systems that can be described by a paradigmatic model like the KS equation, in its different (conserved or non-conserved; deterministic or stochastic) forms, especially when both, chaotic and stochastic fluctuations may be operative at comparable space-time scales.

\begin{acknowledgments}
This work has been supported by Ministerio de Econom\'{\i}a y Competitividad, Agencia Estatal de Investigaci\'on, and Fondo Europeo de Desarrollo Regional (Spain and European Union) through grant No.\ PGC2018-094763-B-I00. E.\ R.-F.\ also acknowledges financial support by Ministerio de Educaci\'on, Cultura y Deporte (Spain) through Formaci\'on del Profesorado Universitario scolarship No.\ FPU16/06304.
\end{acknowledgments}

\newpage

\appendix

\begin{widetext}

\section{Derivation of the stochastic KS equation with conserved noise and additional numerical results on Eqs.\ (3) and (5)}


\bigskip

\subsection{Derivation of the Kuramoto-Sivashinsky equation with conserved noise}

\begin{figure}[b!]
\begin{center}
\includegraphics[width=0.5\columnwidth]{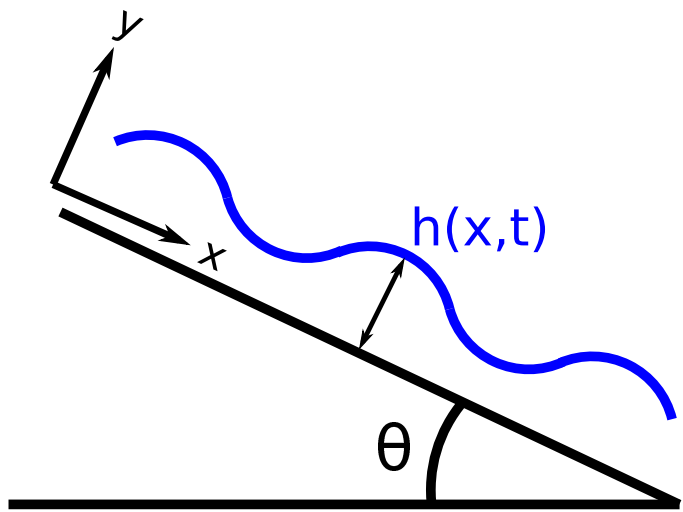}
\caption{\footnotesize Sketch of a thin liquid film falling down a rigid plane inclined at angle $\theta$.}
\label{fig:Plano}
\end{center}
\end{figure}

While the Kuramoto-Sivashinsy (KS) equation has been derived as a physical model in many different contexts, either in the deterministic case or subject to non-conserved noise, there does not seem to be so many analogous explicit derivations in which this equation comes out perturbed by conserved noise. In what follows, we provide one such derivation in the context of falling liquid films, not far from e.g.\ a similar one reported in \cite{Pradas12} for the case of non-conserved noise.

Consider a liquid film which is falling down an inclined plane (see Fig.\ \ref{fig:Plano}) and is so thin that thermal fluctuations can no longer be  neglected \cite{Seeman01,Mecke05}. Assuming fluid incompressibility, the evolution equation for the film thickness, $h(x,t)$, can be obtained from the mass balance $ h_t + \left( \int_0^h u dy \right)_x = 0 $, where $u(x,t)$ is the streamwise ($x$) component of the fluid velocity field, see Fig.\ \ref{fig:Plano} for coordinate conventions. Actually, the full velocity field $(u,v)$ can be obtained from the balance of linear momentum \cite{Pradas12}, namely,
\begin{equation}
    \rho (u_t + u\ u_x + v \ u_y) = \mu (u_{xx} + u_{yy}) - p_x + \rho \ g \ \sin{\theta} + S^{xx}_x + S^{xy}_y ,
\end{equation}
\begin{equation}
    \rho (v_t + u\ v_x + v \ v_y) = \mu (v_{xx} + v_{yy}) - p_y - \rho \ g \ \cos{\theta} + S^{yx}_x + S^{yy}_y ,
\end{equation}
where $\rho$ is the liquid density, $\mu$ is the liquid viscosity (both assumed constant), $p$ denotes hydrostatic pressure, $g$ is the acceleration of gravity, and $S^{ij}$ are the components of a symmetric, zero-mean, delta-correlated fluctuation tensor, $S$. implementing thermal fluctuations in the stress tensor as in classical stochastic hydrodynamics \cite{Mecke05}, and subindices denote partial derivatives.

We consider non-slip, no-penetration boundary conditions $u=v=0$ at the planar rigid substrate ($y=0$) and a simple stress balance at the free surface of the film ($y=h(x,t)$), namely, $ || \vec{n} \ T \ \vec{n} || = \gamma \, \mathcal{C}$ and $ || \vec{n} \ T \ \vec{t} || = 0 $, where
$\mathcal{C}$ is the curvature of the free surface, $\gamma$ is surface tension, assumed isotropic \cite{Pradas12}, the unit normal and tangential vector are
$$ \vec{n} = \frac{1}{\sqrt{1+h_x^2}} \left(
\begin{array}{c}
     - h_x  \\
      1
\end{array}
\right), \qquad
\vec{t} = \frac{1}{\sqrt{1+h_x^2}} \left(
\begin{array}{c}
      1  \\
      h_x
\end{array}
\right),$$
the stress tensor for this Newtonian fluid, including thermal fluctuations, reads
$$
T= \left(
\begin{array}{cc}
    -p-\Pi+2\mu u_x & \mu (u_y+v_x)  \\
     \mu (u_y+v_x) & -p-\Pi + 2 \mu v_y
\end{array}
\right) + \left(
\begin{array}{cc}
    S^{xx} & S^{xy} \\
    S^{yx} & S^{yy}
\end{array}
\right),
$$
and $\Pi$ is the disjoining pressure \cite{Seeman01,Mecke05}.

Now, we consider the average thickness, $h_0$, of the liquid layer as a typical length scale, $w_0=\rho g h_0^2 \sin{\theta}/2\mu$ as a velocity scale, $w_0/h_0$ as a time scale, and $\mu w_0/h_0$ as a representative scale for pressure and stress, and use all these to rewrite the previous equations in dimensionless units. The resulting momentum balance equations become
\begin{equation}
    {\rm Re} \, (u_t + u\ u_x + v \ u_y) = u_{xx} + u_{yy} - p_x + 2 + S^{xx}_x + S^{xy}_y ,
\end{equation}
\begin{equation}
    {\rm Re} \, (v_t + u\ v_x + v \ v_y) = v_{xx} + v_{yy} - p_y - 2 \cot{\theta} + S^{yx}_x + S^{yy}_y ,
\end{equation}
where ${\rm Re}=\rho w_0 h_0/\mu$ is the Reynolds number. The stress balance at the free surface ($y=h$) yields
\begin{equation}
    p+\Pi = \frac{ h_x^2 ( 2 u_x + S^{xx} ) - h_x [2(u_y+v_x)+2S^{xy}] +2 v_y + S^{yy} }{1+h_x^2} - \frac{\gamma}{\mu w_0} h_{xx} ,
\end{equation}
\begin{equation}
    0 = -h_x^2 (u_y+v_x+S^{xy}) + h_x [S^{yy}-S^{xx} + 2 (v_y-u_x)] + u_y+v_x + S^{yx} .
\end{equation}
Now, we introduce a small parameter $\epsilon$ and the new variables $x'=\epsilon x $, $t'=\epsilon t$, and $v'=v/\epsilon$, adapted to a lubrication approximation \cite{Pradas12} within which the cross-stream dimension of the film will be considered much smaller than its streamwise extent. We consider the capillary number, ${\rm Ca}=\mu w_0 / \gamma$, to be order $\epsilon^2$ and define ${\rm Ca}'={\rm Ca}/\epsilon^2$. We expand $u=u_0+\epsilon u_1 + \mathcal{O}(\epsilon^2)$, $v'=v'_0+\epsilon v'_1 + \mathcal{O}(\epsilon^2)$ and $p=p_0+\epsilon p_1 + \mathcal{O}(\epsilon^2)$, and consider $S^{xx}$, $S^{yy}$ to be $\mathcal{O}(\epsilon^{-2})$ and $S^{xy}$, $S^{yx}$ to be $\mathcal{O}(\epsilon^{-1})$ \cite{Mecke05}. Last, by defining ${S^{xx}}'=S^{xx}/\epsilon^2$, ${S^{yy}}'=S^{yy}/\epsilon^2$,
${S^{xy}}'=S^{xy}/\epsilon$, and ${S^{yx}}'=S^{yx}/\epsilon$ the momentum balance equations and the surface boundary conditions become, respectively,
\begin{equation}\label{LMx}
    {\rm Re} \ \epsilon (u_{t'} + u\ u_{x'} + v' \ u_y) = \epsilon^2 u_{x'x'} + u_{yy} - \epsilon p_{x'} + 2 + \epsilon^3 {S^{xx}}'_{x'} + \epsilon {S^{xy}}'_y ,
\end{equation}
\begin{equation}\label{LMy}
    {\rm Re} \ \epsilon^2 (v'_{t'} + u\ v'_{x'} + v' \ v'_y) = \epsilon^3 v'_{x'x'} + \epsilon v'_{yy} - p_y - 2 \cot{\theta} + \epsilon^2 {S^{yx}}'_{x'} + \epsilon^2 {S^{yy}}'_y ,
\end{equation}
\begin{equation}\label{bound1}
    p+ \Pi = \frac{ \epsilon^2 h_{x'}^2 ( 2 \epsilon u_{x'} + \epsilon^2 {S^{xx}}' ) - \epsilon h_{x'} [2(u_y+\epsilon^2 v'_{x'}]+2 \epsilon {S^{xy}}') +2 \epsilon v'_y + \epsilon^2 {S^{yy}}' }{1+\epsilon^2 h_{x'}^2} - \epsilon^2 \frac{\gamma}{\mu w_0} h_{x'x'} ,
\end{equation}
\begin{equation}\label{bound2}
    0 = - \epsilon^2 h_{x'}^2 (u_y+ \epsilon^2 v'_{x'} + \epsilon {S^{xy}}') + \epsilon^2 h_{x'} [\epsilon({S^{yy}}'-{S^{xx}}') + 2 (v'_y-u_{x'})] + u_y + \epsilon^2 v'_{x'} + {S^{yx}}' .
\end{equation}
We can now compute the velocity profile $u=u_0+\epsilon u_1 + \mathcal{O}(\epsilon^2)$. At $\mathcal{O}(1)$, Eq.\ \eqref{LMx} becomes $ u_{0 yy} =- 2 $. As $ u_{0 y} = 0 $ at the fluid surface $y=h$ [leading order of Eq.\ \eqref{bound2}] and $u_0=0$ at the substrate $y=0$, we obtain $u_0=2\left( hy-y^2/2 \right)$. Considering the fluid film to be ultrathin, ${\rm Re} \ll 1$ can be neglected, and Eq.\ \eqref{LMy} at $\mathcal{O}(\epsilon)$ becomes $ u_{1 yy} = p_{0 x'} - {S^{xy}}'_y $. Here we have $ u_{1 y} = - S^{yx} $ [Eq.\ \eqref{bound2} at $\mathcal{O}(\epsilon)$] and $u_1=0$ as boundary conditions at the fluid surface and the substrate, respectively, which allow us to obtain the profile for $ u_{1}= - p_{0 x'} \left( hy-y^2/2 \right) - \int_0^y {S^{yx}}' dy $. The $p_0$ contribution can be obtained from Eq.\ \eqref{LMx} at $\mathcal{O}(1)$, $ p_{0y} = - 2 \cot{\theta} $, with $ p_0 = -\Pi - h_{x'x'}/{\rm Ca}' $ as boundary condition at the fluid surface, obtaining $p_0= 2 \cot{\theta} \ (h-y) - \Pi - h_{x'x'}/{\rm Ca}'$.

Finally, mass balance reads $ h_{t'} + \left( \int_0^h u_0 + \epsilon \ u_1 dy \right)_{x'} = 0 $. Using that $\int_0^h u_0 \ dy = 2h^3/3$ and
\begin{equation}
    \int_0^h u_1 \ dy = - \frac{h^3}{3} p_{0 x'} = - \frac{h^3}{3} \left( 2 \cot{\theta} \ h_{x'} - \Pi_{x'} - \frac{1}{{\rm Ca}'} h_{x'x'x'} \right) + \int_0^y \int_0^{y'} S^{yx} \ dy' \ dy
\end{equation}
yields the evolution equation
\begin{equation}
    h_{t'} + \left( \frac{2}{3}h^3 + \epsilon \frac{h^3}{3} \left( - 2 \cot{\theta} \ h_{x'} + \Pi_{x'} + \frac{1}{{\rm Ca}'} h_{x'x'x'} \right)  + \epsilon \int_0^y \int_0^{y'} S^{yx} \ dy' \ dy  \right)_{x'} = 0.
\end{equation}
Taking $ \int_0^y \int_0^{y'} S^{yx} \ dy' \ dy \simeq (h^3/3)^{1/2} \eta $, where $\eta$ is zero average, Gaussian white noise \cite{Mecke05} and $ \Pi = - \phi' $, where $\phi$ is the interface potential \cite{Mecke05}, we finally obtain
\begin{equation}\label{strongly}
    h_{t'} + \left( \frac{2}{3}h^3 + \epsilon \frac{h^3}{3} \left( - 2 \cot{\theta} \ h_{x'} - \phi'_{x'} + \frac{1}{{\rm Ca}'} h_{x'x'x'} \right)  + \epsilon \sqrt{\frac{h^3}{3}} \eta  \right)_{x'} = 0.
\end{equation}

Finally, a weakly-nonlinear expansion allows us to get the KS equation with conserved noise from Eq.\ \eqref{strongly}. Considering very small fluctuations around the flat film solution, $h=1+\epsilon \tilde{h}$, Eq.\ \eqref{strongly} becomes
\begin{equation}\label{preweakly}
    0= \epsilon \tilde{h}_{t'} + \left( [1+3 \epsilon \tilde{h} + 3 \epsilon^2 \tilde{h}^2 + \mathcal{O}(\epsilon^3) ] \left( \frac{2}{3} + \frac{\epsilon}{3}
P_{x'} \right)
+ \epsilon \sqrt{\frac{1+ \mathcal{O}(\epsilon)}{3}} \eta \right)_{x'},
\end{equation}
where
\begin{equation}
    P_{x'}= \epsilon \left[ - 2 \cot{\theta} \ \tilde{h}_{x'} + \frac{1}{{\rm Ca}'} \tilde{h}_{x'x'x'} - \phi'_{x'}(1+\epsilon \tilde{h}) \right].
\end{equation}
If we linearize $\phi'(1+\epsilon \tilde{h}) \simeq \phi' (1) + \phi''(1) \epsilon \tilde{h} $, expand $\tilde{h}_{x'}^2 = 2 \tilde{h} \tilde{h}_{x'}$, and consider the change of variable $ z=x'-2t' $ and $ \tau = \epsilon t' $ (thus $\partial_{t'}=-3\partial_z + \epsilon \partial_{\tau}$ and $\partial_{x'}=\partial_z$) Eq.\ \eqref{preweakly} becomes
\begin{equation}
    \tilde{h}_{\tau} + 4 \tilde{h} \tilde{h}_z - \frac{2}{3} \cot{\theta} \ \tilde{h}_{zz} + \frac{1}{3{\rm Ca}'} \tilde{h}_{zzzz} + \phi''(1) \tilde{h}_{zz} + \frac{1}{\epsilon} \sqrt{\frac{1}{3}} \eta_z = 0.
\end{equation}
By defining $\kappa_0=1/(3 {\rm Ca}')$, $\nu_0=\phi''(1)- 2 \cot{\theta}/3$, and $\tilde{\eta}=\eta/\epsilon$,
\begin{equation}
    \tilde{h}_{\tau} + \nu_0 \tilde{h}_{zz} + \kappa_0 \tilde{h}_{zzzz} + 4 \tilde{h} \tilde{h}_z + \sqrt{\frac{1}{3}} \tilde{\eta}_z = 0,
\end{equation}
which is a particular case of the stochastic KS equation with conserved noise, Eq.\ (3), after coordinates and fields are renamed as $(z,\tau,\tilde{h},\tilde{\eta}) \to (x,t,u,\eta)$, with $D_0=0$, $\tilde{D}_0=1/\sqrt{3}$, and $\lambda_0=-4$.

\clearpage

\subsection{Further numerical results on the fluctuation statistics of the Kuramoto-Sivashinsky equation}

\subsubsection{Real-space two-point correlation function of the KS equation}

As a complement of the reciprocal-space discussion provided in the main text (MT), here we provide details on the behavior of the two-point correlation function in real space,
$$
C(x,t) = \langle \phi(x_0,t)\phi(x_0+x,t)\rangle-\langle \bar{\phi}(t) \rangle^2 ,
$$
which corresponds to the conserved KS equation, Eq.\ (3), for $\phi(x,t) = u(x,t)$, and to the non-conserved KS equation, Eq.\ (5), if we integrate numerically Eq.\ (3) while taking $\phi(x,t)=h(x,t)=\int_{0}^x u(x',t) \, {\rm d}x'$ and $D_0=0$. Results are shown in Fig\ \ref{fig:CORR2P}. As can be seen, the data collapse in both cases to the expected scaling form, $C(x,t) = t^{2\beta} c(x/t^{1/z})$, with $c(y)\sim {\rm cst.}-y^{2\alpha}$ for $y\ll 1$ and $0$ for $y\gg 1$ \cite{Barabasi95,Krug97}, using the corresponding values of the scaling exponents as determined in the MT. For the $h$ field, collapse is to the exact covariance of the Airy$_1$ process, as expected in the growth regime for 1D KPZ scaling with periodic boundary conditions, see references, e.g.\ in \cite{Takeuchi18}.

\begin{figure}
\begin{center}
\includegraphics[width=1\columnwidth]{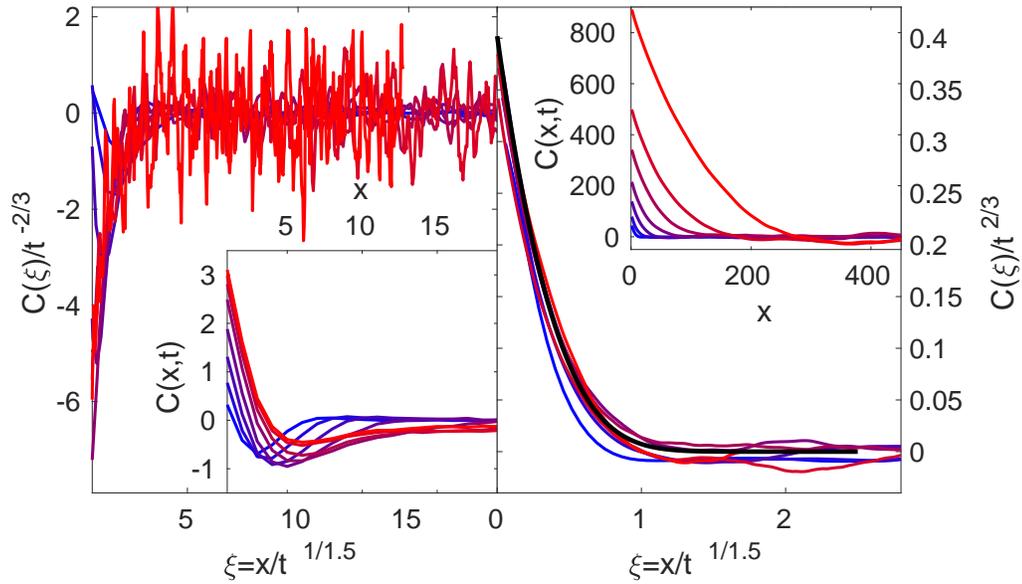}
\caption{{\footnotesize Data collapse of the two-point correlation function $C(x,t)$ at different times in the growth regime (increasing from blue to red), from numerical simulations of Eq.\ (3), for $\phi=u$ (left panel) and $\phi=h=\int_{0}^x u(x',t) \, {\rm d}x'$ (right panel), for $L=2048$, $\delta t = 10^{-2}$, $\nu_0=\kappa_0=1$, $\lambda_0=10$, $\tilde{D}_0=1$, and $D_0=0$. The scaling exponents employed are those determined in the MT for each case. In each panel the inset show the uncollapsed data. The black solid line in the right panel shows the exact covariance of the Airy$_1$ process \cite{Takeuchi18}. All units are arbitrary.}}
\label{fig:CORR2P}
\end{center}
\end{figure}

\clearpage

\subsubsection{Time dependence of the KS fluctuation statistics}

In order to assess in detail the temporal dependence of the fluctuation statistics for the different forms of the Kuramoto-Sivashinsky equation discussed in the MT, here we address the time evolution of both, the skewness and the excess kurtosis of the fluctuations of the $u$ or $h$ fields, see Fig.\ \ref{fig:SerTemp}.
\begin{figure}
\begin{center}
\includegraphics[width=\columnwidth]{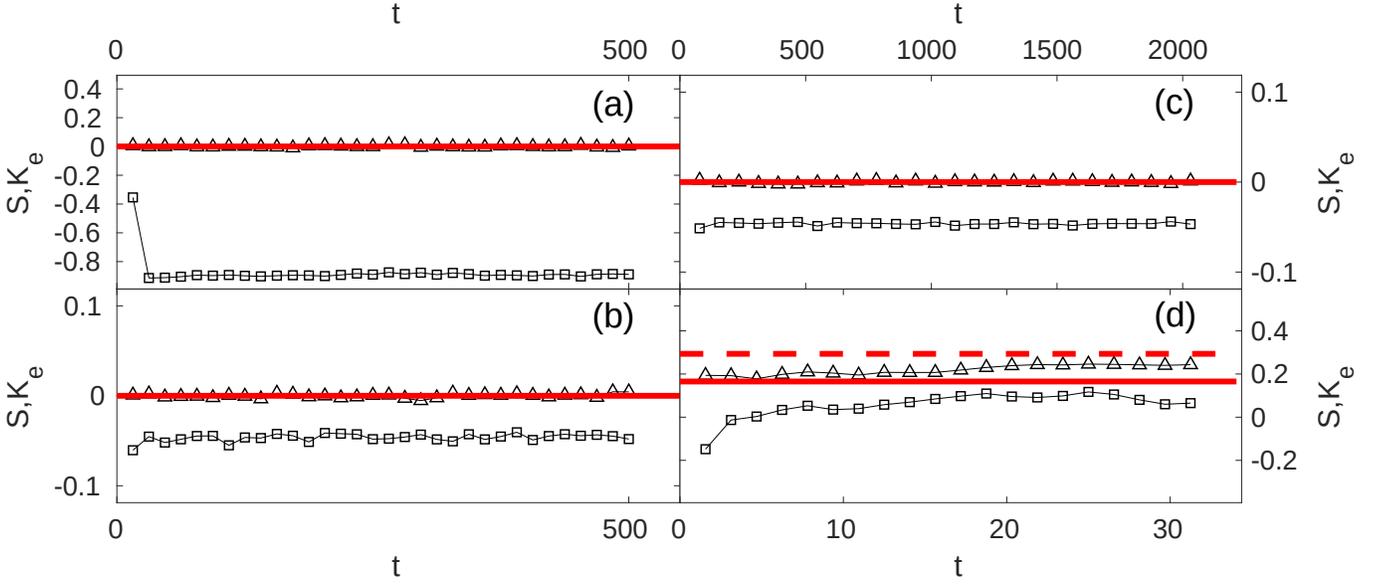}
\caption{\footnotesize Time evolution of the fluctuation skewness $\mathcal{S}$ (triangles) and excess kurtosis $\mathcal{K}_e=\mathcal{K}-3$ (squares) for the conserved Kuramoto-Sivashinsky equation, Eq.\ (3) (using $\nu_0=\kappa_0=1$, $\lambda_0=10$), with (a) $D_0=\tilde{D}_0=0$, (b) $D_0=0$, $\tilde{D}_0=1$, and (c) $D_0=0.1$, $\tilde{D}_0=0$, and for the non-conserved Kuramoto-Sivashinsky equation, Eq.\ (5) (using $\nu_0=\kappa_0=1$, $\lambda_0=10$) with $D_0=1$ (d). Red solid (dashed) lines correspond to the exact skewness (excess kurtosis) value of the Gaussian (a,b,c) and GOE-Tracy-Widom (d) distributions.}
\label{fig:SerTemp}
\end{center}
\end{figure}
Specifically, this figure presents results for: the deterministic KS equation, Eq.\ (3) with $D_0=\tilde{D}_0=0$ [panel (a)], the KS equation with conserved noise, Eq.\ (3) with $D_0=0,\tilde{D}_0\neq0$ [panel (b)], the KS equation with non-conserved noise, Eq.\ (3) with $D_0\neq0, \tilde{D}_0=0$ [panel (c)], and the KS equation for the $h$ field with non-conserved noise, Eq. (5) with $D_0\neq0, \tilde{D}_0=0$ [panel (d)]. In all cases, times are prior to  saturation to steady state.

In general, we can observe that the fluctuation distributions remain largely unchanged over time. Panels (a-c) actually show that, in the corresponding systems (and at variance with the behavior of the 1D KPZ equation), the PDF within the nonlinear time evolution actually coincides with the corresponding PDF at saturation. Such a saturation PDF is reported in \cite{Hayot93} for the deterministic KS equation [panel (a)], while in the cases of the stochastic KS equation with conserved [panel (b)] and non-conserved [panel (c)]  noise results are only available for the corresponding stochastic Burgers equation with which they share asymptotic scaling behavior (in terms of scaling exponents and, presumably, of fluctuation PDF, given that these are stochastics-dominated systems with noise amplitudes above threshold, see Figs.\ 2 and 3), reported in \cite{Rodriguez-Fernandez20} and \cite{Rodriguez-Fernandez19}, respectively.
On the other hand, while the full PDF corresponding to panels (a,b,d) is provided in the MT, this is not the case for the stochastic Kuramoto-Sivashinsky equation with non-conserved noise [panel (c)], whose PDF in the nonlinear regime prior to saturation is presented in Fig.\ \ref{fig:PDF_KSuNC}.

More specifically, the time evolution of the skewness and excess kurtosis shown in Fig.\ \ref{fig:SerTemp} implies a PDF which exhibits a symmetric, notably platykurtic, non Gaussian behavior for the deterministic conserved KS equation [panel (a)], and almost Gaussian behavior in the conserved KS equation with conserved [panel (b)] or non-conserved noise [panel (c)]. Finite-size deviations of the excess kurtosis from their exact zero values seen in panels (b,c) are comparable to similar deviations in the corresponding cases of the stochastic Burgers equation with conserved or non-conserved noise, respectively \cite{Rodriguez-Fernandez20,Rodriguez-Fernandez19}. Finally, the PDF reproduces closely the expected GOE-TW behavior for the non-conserved KS equation with non-conserved noise, Eq.\ (5), see Fig.\ \ref{fig:SerTemp}(d). However, this case is well-known to feature quite different PDF behavior at steady state, as is the case in the 1D KPZ universality class \cite{Takeuchi18}.

\begin{figure}[t!]
\begin{center}
\includegraphics[width=0.75\columnwidth]{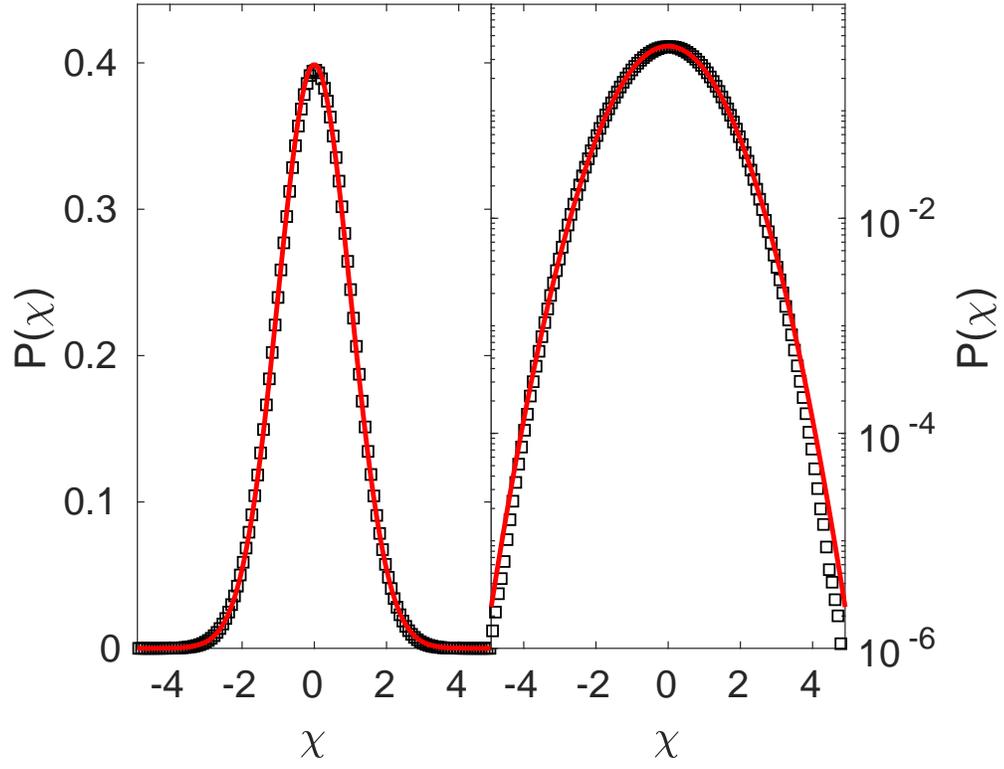}
\caption{{\footnotesize Fluctuation PDF for the Kuramoto-Sivashinsky equation, Eq.\ (3) (using $\nu_0=\kappa_0=1,\lambda_0=10$) with non-conserved noise ($\tilde{D}_0=0$, $D_0=0.1$) in nonlinear regime prior to saturation. The solid line provides the exact Gaussian form. The right panel is a linear-log representation of the same data shown on the left panel.}}
\label{fig:PDF_KSuNC}
\end{center}
\end{figure}

\clearpage

\subsubsection{Transition between chaotic and stochastic PDF}

Figure \ref{fig:TRANS} depicts a more detailed view than that provided by Fig.\ 2, on the transition between the deterministic (chaotic) and the stochastic PDF, which takes place in the KS equation, Eq.\ (3), for increasing values of the conserved-noise amplitude. In the MT, we have identified the threshold value $\tilde{D}_{0,c}\simeq 0.0016$ as that value of $\tilde{D}_0$ above which the kurtosis departs from its deterministic ($\tilde{D}_0=0$) value, see Fig.\ 3. Moreover, for $\tilde{D}_{0}>\tilde{D}_{0,c}$ the scale-dependent Lyapunov exponent $\Lambda(\epsilon)$ changes qualitative behavior from chaos- to stochastic-dominated fluctuations, see Fig.\ 4. In contrast, Fig.\ \ref{fig:TRANS} illustrates how this change is more difficult to see by naked-eye inspection of the form of the field PDF. Note that for all cases considered in this figure, $\tilde{D}_{0}>\tilde{D}_{0,c}$. The two leftmost panels display PDF with inflection points which are akin those characteristic of the PDF for purely chaotic fluctuations ($\tilde{D}_0=0$). Nevertheless, such inflection points disappear once the stochastic-noise amplitude increases even further above $\tilde{D}_{0,c}$, beyond which the fluctuation PDF eventually reaches fully-Gaussian form, see Fig.\ 2.

\begin{figure}
\begin{center}
\includegraphics[width=1\columnwidth]{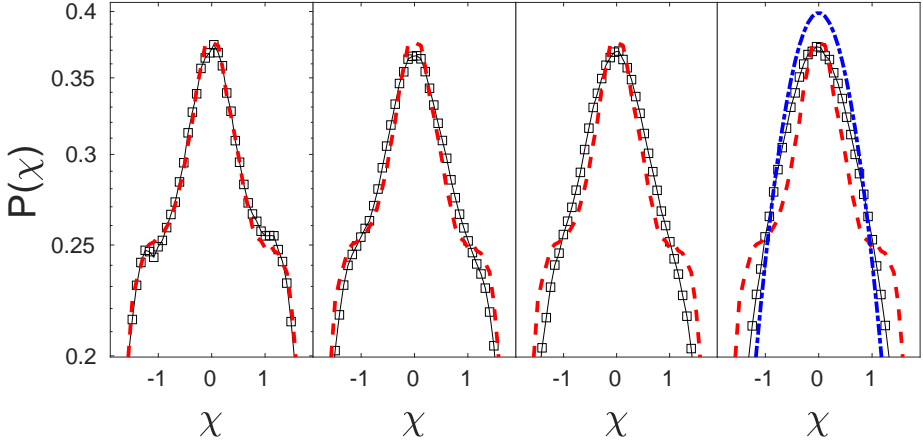}
\caption{{\footnotesize Fluctuation PDF for the standardised $u$-fluctuations ($\chi$) of the Kuramoto-Sivashinsky equation, Eq.\ (3), using $L=2048$, $\delta t = 10^{-2}$, $\delta x = 1$, $\nu_0=\kappa_0=1$, $\lambda_0=10$, $D_0=0$, and increasing values of $\tilde{D}_0= 0.0063, \ 0.016, \ 0.025$, and $0.040$, left to right. Red dashed lines show the PDF for the deterministic $\tilde{D}_0=0$ case; black solid lines guide the eye for the numerical values shown as squares; the blue dotted-dashed line in the rightmost panel shows the exact Gaussian PDF. All vertical axes are in logarithmic scale. Averages are made over $10$ noise realizations.}}
\label{fig:TRANS}
\end{center}
\end{figure}

\end{widetext}

\end{document}